# Exploring Small-World Network with an Elite-Clique: Bringing Embeddedness Theory into the Dynamic Evolution of a Venture Capital Network

Weiwei Gu[1], Jar-der Luo[*,1], Jifan Liu

Department of Sociology, Tsinghua University, Beijing, 100084, P. R. China

## Abstract

This paper uses a network dynamics model to explain the formation of a small-world network with an elite-clique. This network is a small-world network with an elite-clique at its center in which elites are also the centers of many small groups. These leaders also act as bridges between different small groups. Network dynamics are an important research topic due to their ability to explain the evolution of network structures. In this paper, a Chinese Venture Capital (VC) network was coded from joint investments between VC firms and then analyzed to uncover its network properties and factors that influence its evolution. We first built a random graph model to control for factors such as network scale, network growth, investment frequency and syndication tendency. Then we added a partner-selection mechanism and used two theories to analyze the formation of network structure: relational embeddedness and structural embeddedness. After that, we ran simulations and compared the three models with the actual Chinese VC network. To do this we computed the elite-clique's EI index, degree distribution, clustering coefficient distribution and motifs. Results show that adding embeddedness theories significantly improved the network dynamic model's predictive power, and help us uncover the mechanisms that affect the formation of a small-world industrial network with an elite-clique at its center.

*Keywords:* Venture capital industry, Network dynamics, Embeddedness, Syndication network

## Introduction

In this paper, we build a partner-selection mechanism into an industrial network model, and fit it to the Chinese Venture Capital (VC) industry. In the Chinese VC industry there are major players (known as 'leaders' or 'elites' in this paper) that lead a group of smaller companies (known as 'followers' or 'satellites' in this paper) to form small groups with center-satellites structure. These major companies also act as bridges, by connecting to other small groups. In other words, the VC network is a special type of small-world (Watts & Strogatz, 1998), in which the central roles of small groups are also bridges between these groups. We use the term center-satellites group to refer to small, comparatively isolated "caves" (a term indicates isolated groups) in a small-world network. This paper will use this term to specify a leader that is surrounded by a small group of followers. Major players frequently cooperate with each other, and through their cooperation they form an elite-network (Useem, 1984) which is at the center of the whole industry. For this reason, we call these major players industry "elites".

---

[*] Corresponding author.
   E-mail address: jdluo@mail.tsinghua.edu.cn (Jar-der. Luo)
[1] These authors share the first co-authorship.

This paper will present a partner-selection mechanism that integrates embeddedness theory to show the formation of a small-world network with an elite-clique at its center. These elites are the centers of their respective center-satellites groups and also act as bridges to other small groups. In an industry of this type, a major firm often initiates a business plan and then chooses followers to join in the plan. For example, in a VC investment plan, there is always a major investor who assesses an investment opportunity, is responsible for setting up the plan, and organizes the syndication partners. There may be one seat in an investee's board for investors, and in general, this major player will represent investors by being on the board. This paper incorporates the selection mechanism stated above into the model, and illustrates significant structural results of this selection mechanism in the dynamics of the industry network.

Modeling networks serves at least two purposes. Firstly, it helps us understand how networks form and evolve, e.g., in terms of community evolution, changes in network statistics, and emergence of central roles, etc. Secondly, in studying simulations of network-dependent social processes, successful network models can be used to specify the structure of interaction, such as link prediction (Tylenda, Angelova, & Bedathur, 2009) and community detection (Nguyen, Dinh, Tokala, & Thai, 2011). A large variety of models have appeared in the physics-oriented network literature in recent years. Some models focused on understanding the structure and evolution of these large-scale networks (Barabási et al., 2002; Garlaschelli & Loffredo, 2005; Rosenkopf & Padula, 2008), while other models shed light on the high level of growth and topological characteristics of these networks. For example, Viswanath, Mislove, Cha, & Gummadi (2009) presented the interaction among users in the Facebook online social network, and they found that a minority of the user pairs generate a majority of the activity. The models mentioned above can be classified into two main categories: 1) Models where new links may be added depending on the local network structure. 2) Models where the probability of establishing or removing a link would depend solely on the nodal attributes.

In this paper, we will present the structure and evolution of a VC network by combining properties based on nodal attributes and local network properties. Nodal properties include investment frequency and syndication tendency for each node. The partner-selection mechanism and embeddedness theory, which includes relational and structural embeddedness (Granovetter, 2017), are brought in to incorporate network structure into the model. VC firms invest in start-up companies that have potential, but lack capital. This involves dealing with certain risks and uncertainties. As a way of sharing resources and risks, syndication between VC companies is a common investment strategy to get more opportunities and reduce risk.

We introduce the partner-selection mechanism to help us explore what shapes this network. Through this mechanism, we show how a major player in the VC industry selects a partner for an investment. This mechanism first incorporates findings from relational embeddedness theory on partner selection as well. These findings are primarily about inviting frequent syndication partners (VC firms which have multiple joint investments) to jointly invest in a company. Our other model then adds structural embeddedness (to relational embeddedness), which broadens the scope of joint investments from friends to friends' friends. In other words, friends of friends are included in the syndication network of future investments. Finally, we compare the different models with the Random Joint Investment model, in which VC firms randomly choose syndication partners. In the models, we control for network size, growth, investment frequency and syndication tendency. These are calculated from real data. We also select degree, tie strength and clustering coefficient distribution as macro-level indicators and some motifs as micro-level indicators to compare the three models with real network data. Our comparison shows that both the relational and structural embeddedness models fit the real network well.

In the following sections, we will introduce the network features of the Chinese VC industry, as well as syndication partner-selection mechanisms, and the relevant literature. After that, our data collection process is described. We then build the three network dynamics models. At last, this paper will compare the models and simulate the structural embeddedness model to show the formation of the industrial network structure.

## Theoretical Background

### The Evolution of a VC Industrial Network

Syndication among venture capital firms is defined as a joint investment by multiple investors in one investee at the same time. Networks are dynamic, new roles are added in and new ties are built, while old ties might be maintained or cut off (Ahuja et al., 2011; Beckman, Haunschild, & Phillips, 2004; Loftus et al., 2005). A VC firm may build and cut ties for many reasons, including: A better network position (Bygrave, 1987); To exercise greater influence on investment partners and target newly-emerging companies (Sorenson & Stuart, 2008); To obtain a higher probability of survival (Hochberg, Ljungqvist, & Yang, 2010). Furthermore, this branch of research investigates the motivations behind networking behaviors, such as diverse geographical and industrial distribution (Kogut, Urso, & Walker, 2007), and multiple connectivity and trend following in the bio-tech industry (Powell, White, Koput, & Owen-Smith, 2005).

There is a relative paucity in the corporate finance literature due to the difficulties of analyzing syndication patterns empirically and verifying the complex motives behind syndication (Lerner, 1994). Attachment patterns and the logic behind relationship building might be the key to better explaining this field (Hochberg et al., 2010). Especially in the context of emerging countries such as China, the difficulties of data collection often impede research. Although Chinese VC firms have many things in common with their western counterparts, they also have differences. One difference in particular is the institutional and cultural emphasis on guanxi and networking (Burt & Burzynska, 2017), which can bring about different activities and relation-building logic (Bruton & Ahlstrom, 2003; Bruton, Fried, & Manigart, 2005).

The Chinese VC industry is characterized by government intervention and newly established, constantly changing institutions. In this kind of high-uncertainty environment, players seek alliances to hedge against various types of risks. Building guanxi, defined as "a high-trust relationship which is relatively independent of the social structure around it" (Burt & Burzynska, 2017; Burt & Opper, 2017), is a common strategy for players in this industry. Major players often mobilize their reputation and resources to attract loyal followers to form guanxi circles, i.e. ego-centered networks with various layers of guanxi, each with different levels of trust and intimacy (Luo et al., 2017). These small circles constitute center-satellites groups in a small-world network. In addition to small groups, a small-world network has a few bridges that connect otherwise isolated groups, so that the cluster coefficients in local networks are very high but the average distance among nodes in a larger network is short (Watts & Strogatz, 1998). In the Chinese VC network, a central actor (elite) generally plays the role of a long-distance bridge, while its followers form a dense in-group network that rarely connects to outsiders (Luo, Ke, Yang, Guo, & Zou, 2018). Furthermore, these major players, or "elites" (Useem, 1984), frequently cooperate with each other, so an elite-clique emerges as the network's center (Lou & Li, 2017). In other words, guanxi makes small groups' leaders and their followers cooperate frequently, and bonds the leaders together to form an elite-clique (Wu, Wu, & Rui, 2009). In the following, we will use the term 'small groups' to indicate center-satellites groups, and 'clique' to specify the elite-clique.

---

**Fig. 1** is about here

---

We use a network model to explain the evolution of the network structure in this sort of industry. This model uses two theories, relational embeddedness and structural embeddedness (Granovetter, 1985), to analyze network dynamics in the Chinese VC industry.

**Relational embeddedness theory**

Long-term cooperation breeds trust (Gulati & Westphal, 1999), and the expectation for future runs of a game ensures that cooperative relations continue (Axelrod & Reisine, 1984; Hardin, 2001). Chinese people tend to cooperate with familiar persons by means of long-term favor exchanges (Hwang, 1987). This suggests that a Chinese VC will initiate a new joint investment with people who have had a cooperative experience with the centered ego (generally the leader or leaders of the VC firm) in the past. Frequent cooperation and strong trust often come from combined motivations, i.e., motivations that mix instrumental and friendship goals (Granovetter, 2017), which in turn promote further syndication. A successful investor always has familiar partners, who form the hard core of this person's ego-centered co-investment network or the centered ego's guanxi circle (Luo, et, al., 2018). Companies who have frequent cooperation, strong trust, and friendship with the centered ego are more likely to be chosen in a new run of syndication, so we include this statement in the model: Frequent cooperation increases the possibility of participation in a new joint investment.

**Structural embeddedness theory**

Simply having strong ties in a dense ego-centered network is not enough to build up a system of relationship developing mechanisms. For example, a new run of cooperation may involve a huge amount of profit or loss, and this may damage relations between old partners. For the sake of maintaining a friendship, the centered ego sometimes avoids inviting familiar partners to join a new investment, and instead opts to recruit an indirect tie as a new partner.

Alternatively, a focal VC firm needs to search for new investment opportunities. A small circle of frequent syndication partners cannot provide these opportunities, so it needs new partners to broaden its search scope. Embeddedness theory argues that good economic decisions depend on whether information comes to you (Granovetter, 2017). Without social relations in the Chinese VC industry, it is hard to get enough information to analyze a project. Relation building is thus needed to access good projects. In other words, one focal actor needs to take continuous action to build relations to access new investment opportunities. Information providers can be weak ties or familiar partners. According to structural-hole theory (Burt, 1992), familiar partners form a dense, small-scale network, while weak ties form a sparse, large-scale network, which tends to be full of non-redundant information. In addition, many holes in this sparse network grant the hole-spanning ego information and control benefits. The sparse network of occasional syndication ties forms a peripheral belt outside the core of a centered ego's guanxi circle. This belt has two functions: The first is to act as a process of trial and error through which unfamiliar partners are tested to see if they are suitable for long-term cooperation. The second function is that weak ties in this peripheral belt may help introduce indirect ties, who may then be involved in syndication projects.

To summarize the above, social ties are important as they enable access to valuable projects and introduce new partners to a centered ego. However, in such a highly uncertain environment, strangers or those with high relational distance are not a good choice to be an investment partner, since information asymmetry makes them untrustworthy. So those with a short path to the centered ego, such as direct ties and friends of friends are chosen. This leads to an additional function: Relational distance is associated with the possibility of joint investment.

Most Chinese VC firms avoid cooperating with strangers, as contracts and contract law are not reliable. Third-party trust (Burt & Knez, 1996) plays an important role in this context. A mediator can transfer his or her guanxi to another friend, while a bridge can provide much needed information at the right time (Burt, 2000). Without knowing detailed information about a possible new partner, a centered ego may find it hard to make a decision. This highlights the second key function of guanxi: To introduce new partners to old friends. In other words, common neighbors help introduce possible syndication opportunities to potential partners. The more common neighbors two VC firms have, the greater the possibility they will become partners. We thus add a third function to our model: The number of common neighbors is positively associated with the possibility of joint investment.

## Methodology

### A Small-World Network with an Elite-clique

In China, major venture capital databases such as China Venture, Zero2IPO and Venture Capital Research Institute's annual reports release data about all public investments. Among them, SiMuTon Database, built by research institute Zero2IPO, collects investment data from the internet and other sources. It includes the investee's name, investment date, and investment stage (initial, expansion, and seed stage). Based on this widely-used database, we identified syndication ties among VCs and built a co-investment network. **Fig. 2** shows how this undirected and weighted network is formed: A tie is created between two VC firms when they invest in the same target firm at the same time. We then collected more investment information from public sources, to compensate for any missing network data. In total, our model includes information for 7,640 investment events and 1,436 nodes (VC firms) and 4,623 edges (co-investment ties among those firms) from 2000 to 2013.

---

**Fig. 2** is about here

---

In **Fig. 3**, we first compute the distance between two VCs. If they are neighbors, then we assign them a distance of one. If they have a common neighbor, then the assigned distance is two. If they are connected by three steps, then the distance is three, etc. If they made eight joint investments in the past, then the distance is 1/8, 7 joint investment is 1/7, etc. The vertical axis indicates the probability of syndication in the next period. **Fig. 3** shows that firms with a distance of three have no chance to cooperate. Also, the probability of syndication increases as the number of past-joint investments between two VCs increases.

---
**Fig. 3** is about here
---

We compute our network structure through several steps. In the first step, this study will find the major players (elites) in the VC network. We apply the Graph Attention algorithm (Velickovic et al., 2017) to identify these elites. The Graph Attention algorithm uses VC nodes and their first-order neighbors as its input data. We then use the Adam SGD optimizer (Kingma & Ba, 2015), which computes the attention coefficient in each pair of nodes. The attention coefficient indicates a node's neighbors' influence on the partnered node. Based on the assumption that the more attention a node obtains, the more important it is, (Gu, Gong, Lou, & Zhang, 2017), we sum each node's total attention and then rank all VC nodes in descending order. The Graph Attention method takes the attention coefficient matrix of each node-pair, a 1,436 times 1,436 matrix, and identifies out the top 42 VC firms, which we define as industry elites.[2]

Table 1 shows the average degree, investment frequency, k-shell and betweenness for elites and their satellites. Elites have a higher investment frequency and a larger amount of co-investments with other VCs in comparison to satellites. Their average k-shell is also high, meaning they have more connections with other important players. In addition, these elites' average betweenness centrality is much larger than the followers', telling us that they act as bridges in the whole network.

We also compute the density within the elite-clique as well as in each center-satellites group. The elite clique has tight within-group interaction, so the density is much higher than average, and the EI index (within-group density divided by whole-network density) is as high as 123. This indicator shows that the elites cooperate so frequently that they form a small clique at the center of the whole industrial network. We then calculate the center-satellites group density around each elite and average the results. The average center-satellites group density is also high in comparison to the whole network density. The EI index is 25, showing that followers in a center-satellites group tend to cooperate with others within a   group, rather than with other   groups' members.

From the analyses stated above, we outline network properties that indicate the network structure of the Chinese VC industry, i.e. a small-world network with some elites (major players), each of whom leads a small group of satellites. These elites are also bridges among center-satellites groups and bond together so that an elite-clique emerges at the center of the entire network. Finally, we compare these indicators with our models' predictions to show the models' explanatory power.

---
Table 1 is about here
---

Next, we look at the specifications and parameters of the three models. The first model integrates relational embeddedness, the idea that frequent syndication partners tend to jointly invest in a start-up. The second model, structural embeddedness, adds an additional function to the relational embeddedness model: The probability of a VC firm to invite friends of friends to join in new investments. We compare the relational and structural embeddedness models with a Random Joint Investment model while controlling for network scale, network growth, investment frequency and syndication tendency. Below are the specifications and parameters of the three models.

---

[2] As a rough test of the validity of this computing method, we used the Delphi method to identify "elites" in the Chinese VC industry. Elites are defined as companies that can find good investment opportunities, play a leading role in investment activities, set up investment plans and organize their followers to be syndication partners. In other words, they are center-satellites group leaders. We then explained our definition of elites to four experts, including one chairman of a national leading research institute, two CEOs of big domestic VCs, and one CEO of a big foreign investor. They eventually provided us with a list of 42 elite VCs, and the others are listed as followers of these leaders. We tested the validity of the Graph Attention algorithm with the real VC network. It achieved a prediction accuracy of 0.78 and F1 Score = 0.80 in finding the 42 elites.

**Random Joint Investment Model**

We focus on the changes from 2000 to 2013, so the models simplify each year to a moment of t and control network size by setting the evolution iteration $t = 14$. The growth rate of the number of venture capital institutions is taken as a function of time. We then apply the exponential function (1) to fit the VC growth rate with the starting number of VCs $m_1 = 75$. The growth function of the number of investors is expressed as:

$$m_t = f_A(m_{t-1}) = 75 \times 1.3^t \tag{1}$$

Based on industry experience (Jovanovic, 1982), we give a rough linear estimation formula for the target firms (investees), with the starting number of firms $n_1 = 375$. Function 2 shows how targeted firms grow as a function of VC firms.

$$n_t = 5 \times (m_t) \tag{2}$$

Different VC firms have different investment frequencies. Large VC firms usually invest frequently, while small VC firms tend to have a low investment frequency (Hochberg, Ljungqvist, & Yang, 2007), as shown in **Fig. 4**.

**Fig. 4** is about here

In order to describe this property and control for network density, we assign a time-invariant variable of investment frequency (Fi) to each VC firm. This property defines how many firms a VC will invest in each time. We apply a stair-case function to curve the investment frequency property. To simplify this property, VC firms are divided into three categories, as shown in Table 2. The first third have a low frequency (Fi = 0.26), the other third a middle frequency (Fi = 0.796), and the final third have a high frequency (Fi = 5.047). All these statistics are computed from real data.

Table 2 is about here

This property is static, which means once a new VC firm is added into the evolution process, its investment frequency won't change in different iterations. Whenever two VC firms randomly invest in the same firm, the Random Joint Investment model counts one co-investment between them.

Table 2 shows the control variables for the model and the evolution of network size.

**Relational Embeddedness Model**

Relational embeddedness refers to trust created by repeated cooperation between a pair of VC firms. In order to describe this property, we divide evolution process in each time t into two stages: random joint investment and inviting cooperation. The first stage is the same as the Random Joint Investment model, i.e. VC firms randomly choose target firms. In the second stage, VC firms invite frequent syndication partners to co-invest in target firms.

**Fig. 5** is about here

Different VC firms have different syndication tendencies (Qi), i.e. the number of partners a VC will invite in each run of investment. According to the syndication tendency distribution of real data shown in **Fig. 5**, we classify the syndication tendency of network nodes into three categories--the first one-third have a high tendency (Qi = 0.90), the second third is medium (Qi = 0.59) while the last one is very low (Qi = 0.20). The syndication tendency characterizes the VC firms' willingness to syndicate.

**Table 3** is about here

The invitation strategy also differs between syndication partners. VC firms that operate according to guanxi-orientated behavioral norms are more likely to syndicate with frequent cooperation partners. Table 4 shows that a Chinese VC may invite its closest partners to invest jointly 3 times in every 4 investments. It also shows that if two VC firms have partnered together many times, then the probability they will invite each other in the next run is higher, and the invitation probability grows almost linearly with the number of joint investments. Thus, we measure invitation probability by co-investment frequency.

In this model, we use $n_{ij}$ to represent the co-investment frequency between VC $i$ and $j$ and invitation probability $p_{i,j}$ is defined as:

$$p_{i,j} = \begin{cases} 0 & \text{(no tie between } i \text{ and } j \text{)} \\ \dfrac{n_{i,j}}{\sum_j n_{i,j}} & (n_{i,j} \text{ is the joint investment times)} \end{cases} \qquad (3)$$

Table 4 is about here

In the relational embeddedness model, we control for network size, network growth, investment frequency, syndication tendency and invitation probability. As shown in Table 3 and , there are nine types of VC nodes in the relational embeddedness model. **Fig. 6** shows how a network is formed according to this model.

**Fig. 6** is about here

**Structural Embeddedness Model**

Firms can use third-party trust to broaden the scope for new investments. Thus, in the structural embeddedness model, we add the probability for VC firms to invite friends of friends to syndicate. The only difference between relational embeddedness and structural embeddedness lies in the invitation strategy. In the relational embeddedness model, if there is no direct tie between two VC nodes, there is no chance for them to invite each other during the second stage in any time stamp. But in the structural embeddedness model, whenever two VC firms are linked indirectly by common neighbors, as shown in **Fig. 7**, there is still some probability of invitation. The structural embeddedness model also has two stages, random investment and inviting cooperation stages. $n_{ij}$ denotes the co-invest time. N is the co-invest matrix. We apply M = N * N to denote the second-order relationship between node pairs. The invitation probability $p_{i,j}$ is defined in Function 4:

$$p_{i,j} = \begin{cases} \dfrac{m_{i,j}}{\max(M)} & \text{(no tie between } i \text{ and } j \text{)} \\ \dfrac{n_{i,j}}{\sum_j n_{i,j}} & (n_{i,j} \text{ is the syndication frequency)} \end{cases} \qquad (4)$$

## Analytical Results

To measure the predictive power of the three models, we compare the models' simulation results with actual network data. We check macro-level fitness through the distribution of degree, cluster coefficient and tie strength. Micro-level fitness is tested by computing the number of various types of motifs.

### Testing by Macro-level Distribution

We selected several macro-level network indicators such as degree distribution, tie-strength distribution and clustering-coefficient distribution to compare the Random Joint Investment, relational embeddedness and structural embeddedness models with the real industrial network. **Fig. 8**, **9** and **10**, show that the relational embeddedness and structural embeddedness models are far better than the Random Joint Investment model at predicting macro-level indicators computed from real data. In addition, the structural embeddedness model does even better than the relational embeddedness model.

### Testing by Micro-level Motif

Here we introduce various types of motifs, which are microscopic indicators in network analysis. Motifs are patterns (sub-graphs) that recur within a network much more often than expected in a random graph. Finding motifs can not only help us uncover the structure of a complex network but also provide a deep insight into its formation mechanisms. In Table 5, we compare the number of motifs generated through simulation results with those of the real network. Table 5 shows that the structural and relational embeddedness models perform much better than the Random Joint Investment model. In the real-world network, structural and relational embeddedness models have no quadrangle motifs. This means that if two VC firms have two common neighbors, there will definitely be a joint investment between them in the near future. We also find that the structural embeddedness model is better than the relational embeddedness model, which proves that inviting friends' friends to participate in joint investments is an important mechanism in the Chinese VC industry.

### The Predictive Power of the Structural Embeddedness Model

In the final stage, the structural embeddedness model is used to simulate the evolutionary process of the Chinese VC industry network, and predict the properties of a small-world network with an elite-clique at its center. As shown by the betweenness, degree, elite-clique EI index and average center-satellites groups' EI index in Table 1, the structural embeddedness model has far better predictive power than the Random Joint Investment model. Betweenness and degree simulation results show that the leaders of center-satellites groups are also bridges in the whole industrial network. The elite-clique EI index is very similar to the statistics computed from real data, in that the density within the elite clique is extremely high. The average center-satellites groups' EI index is even higher than that computed from real data, and reveals that the connections within each small group are on average much denser than those between groups.

To evaluate the structural embeddedness model's predictive power, we compute the eight network property indicators mentioned in Table 1 for the whole time frame of our study. We compare them with real data in each year by calculating the correlation among these indicators. **Fig. 11** shows that the structural embeddedness model's correlations with the real network grow during network evolution. That means, the small-world network with an elite-clique, in which center-satellites group leaders play the bridging roles at the same time, gradually forms in the evolution process, as shown in the simulation of the structural embeddedness model.

**Fig. 11** is about here

## Conclusion and Discussion

### Exploring the VC Industry Network Structure

Our paper tries to explore the formation of a small-world industrial network with an elite-clique at its center. According to the structural embeddedness model in the VC industry, there is a group of major players who often invest together. They seek out followers to join them at the same time. These major players invite partner based on a selection principle, which is that frequent syndication partners are more likely to be asked to participate in future joint investments, as suggested by relational embeddedness theory. In some cases strangers may be invited to participate in joint investments. Structural embeddedness theory posits that friends' friends with a larger number of common neighbors are more likely to be invited.

**Fig. 11** shows that as the structural embeddedness model evolves, its correlation with real-world data increases. In other words, the partner-selection mechanism integrated into the model fits the real network well. That means, major players initiate most joint investments, and select familiar syndication partners to participate in these investments. Gradually, followers bond together around one major player and a center-satellites group forms. Structural embeddedness states that this major player also seeks partners outside of its small group, and companies with more common neighbors are more likely to be accepted as partners. Other major players have much more connections in the industrial network than ordinary VC firms. Therefore, they have more common neighbors with other major players, and more chances to become direct cooperation partners. In the long-term, these major players build dense ties with each other and an elite-clique emerges at the center of the whole network.

Based on our results we found that the structural embeddedness model is better than the relational embeddedness model at predicting real-world data. Both the structural embeddedness and relational embeddedness models are much better than the Random Joint Investment model. We control for network scale, network growth and investment frequency which, even though they are network statistics computed from real data, are not enough to make a good prediction. In the statistical distribution for both macro-level network features and micro-level motif numbers, both the embeddedness models have much better prediction results than the base model. Social embeddedness theory is the foundation for the relational embeddedness and structural embeddedness models. Both relational and structural embeddedness theories are confirmed as important factors in network dynamics (Burt, 2010; Gulati, 1999), and in particular in the VC field (Powell et al., 2005). That is, existing cooperation has an important impact on new runs of cooperation, and neighboring positions in the network structure increase the possibility of becoming partners in the future.

Incorporating these theories into a model is very meaningful. They not only got confirmation from the regression analysis, but partner-selection mechanisms can be explored via a network dynamics model. Change in a social system is the result of the interplay between social networks and behaviors (Granovetter, 1973; Holbrook, 2002; Padgett & Powell, 2012), network dynamics is vital for us to understand the evolutionary process of a social system. We find that only building existing network statistics into a dynamic model is not enough to predict a real network. Social theory can help us explore the operating mechanisms that drive change in a network.

## The Usefulness of VC Network Dynamics

The Chinese VC industry is a social system with its own network features and dynamics. Different kinds of social systems may have various mechanisms that affect network dynamics in different ways, so we must carefully examine the behaviors behind these mechanisms. For example, why does a Chinese VC firm only choose indirect ties with common neighbors as potential partners? How do Chinese VCs work with frequent partners? Are there any theoretical findings which may guide us in choosing which model parameters to include? Doing further qualitative studies will help answer these questions. Combining qualitative and quantitative studies reveals more about the processes that shape a network (Small & Newman, 2001). In future studies, we need to do fieldwork to collect qualitative information about how social embeddedness works in the Chinese VC industry. For example, we discovered that there is no closed quadrangles in the real network, structural and relational embeddedness models. That means, whenever two VC firms have two common neighbors, they will become cooperating partners. Why is that? This is a puzzle which deserves our attention in the next run of studies, which may help us uncover more mechanisms behind the evolution process.

Like any study, we still have room for improvement. There are many other network factors not in this model, such as similarity, the tendency to follow trends (Powell et al., 2005), preferential attachment (Barabàsi, 2005; Barabàsi & Albert, 1999) and some institutional explanations, e.g. norms that discourage satellites from becoming bridges (Xiao & Tsui, 2007). Adding these factors into a model will provide us with an even better fit for the actual network. This is especially true in different institutional environments; For example, foreign investors and state-owned VCs face different institutional environments in China, and have different syndication behaviors (Luo, et, al, 2018). These two types of VCs are even organized into two comparatively separate communities (Wang, Zhou, Tang, & Luo, 2016). How do their networks evolve? Do their network structures have significantly different outcomes? Do they have different model parameters? Or, do we need to build different models to explain their varying evolutions? These questions pose interesting challenges for future studies.

Our model sheds light on the Chinese VC industry, but can we generalize it to other industries and those in other cultures? This model should have some explanatory power for all industrial network that have a similar structure to the one shown in **Fig. 1**. However, in addition to the generalized model, social theories provide useful explanations in certain social contexts. It is good to explore the causal mechanisms proposed by social theories qualitatively and quantitatively in different social contexts. Mechanisms that gain confirmation can be built into model to simulate the evolutionary process of a network. Any inconsistency in the model provides us with clues to investigate other mechanisms behind the dynamic process. So, in the next run of studies, new theories may be brought in, and a new model can be built. A good starting point would be testing the embeddedness model with the data of other industry networks. When compared with different cultural or industrial contexts, we may get variation in our model parameters, or even need to modify our model.

This paper is an attempt to explore one particular type of network. More social factors and mechanisms need to be introduced in future studies, and more cases that encompass different social contexts need to be collected.

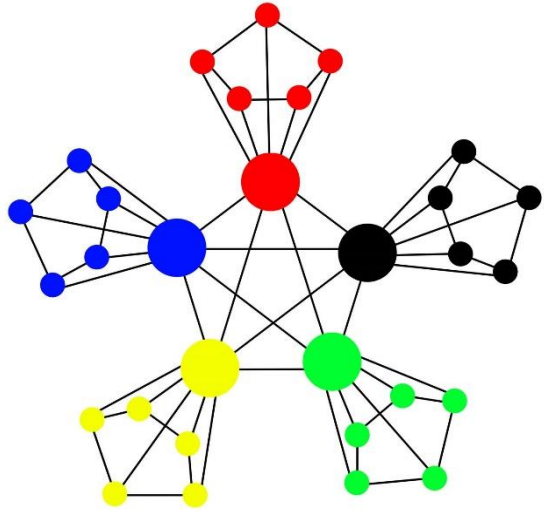

**Fig. 1.** An Example Illustrating the Small-World Network: A small-world network with an elite-clique. Elites are not only the leaders of center-satellites groups but also bridges among these groups.

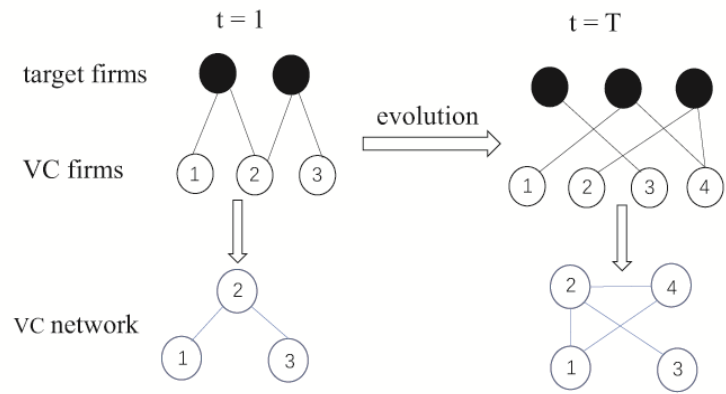

**Fig. 2.** The Formation of Syndication Ties: A tie is created between two VC firms when they invest in the same target firm at the same time.

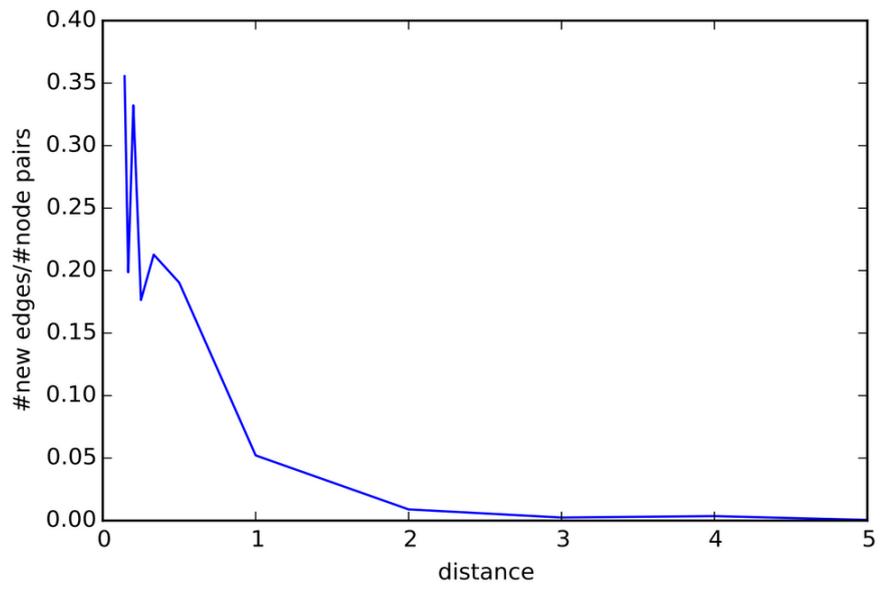

**Fig. 3.** Probability of Syndication and Distance

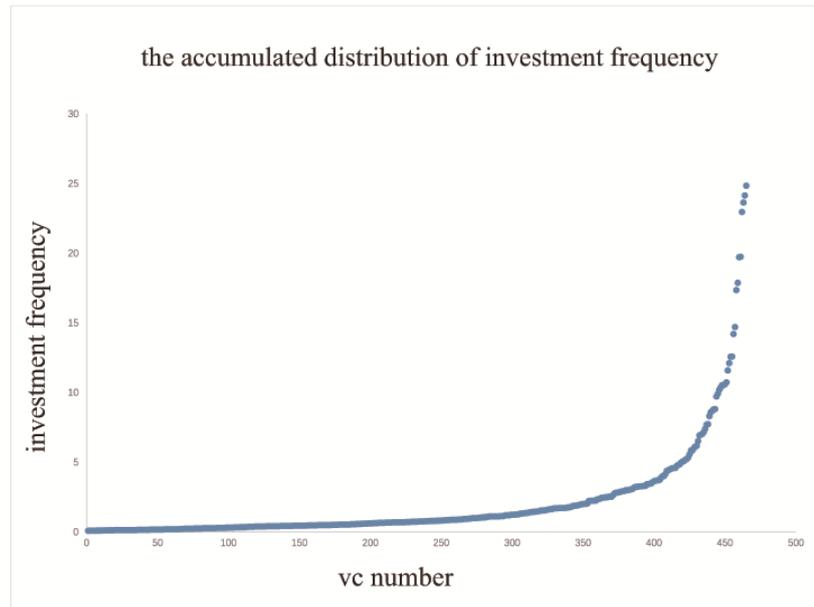

**Fig. 4.** Ranking VC Firms According to Investment Frequency: The X-axis is investment frequency, and the Y-axis is the accumulative number of VC firms. A large number of VC firms have a low investment frequency, while only few investors' frequency are high.

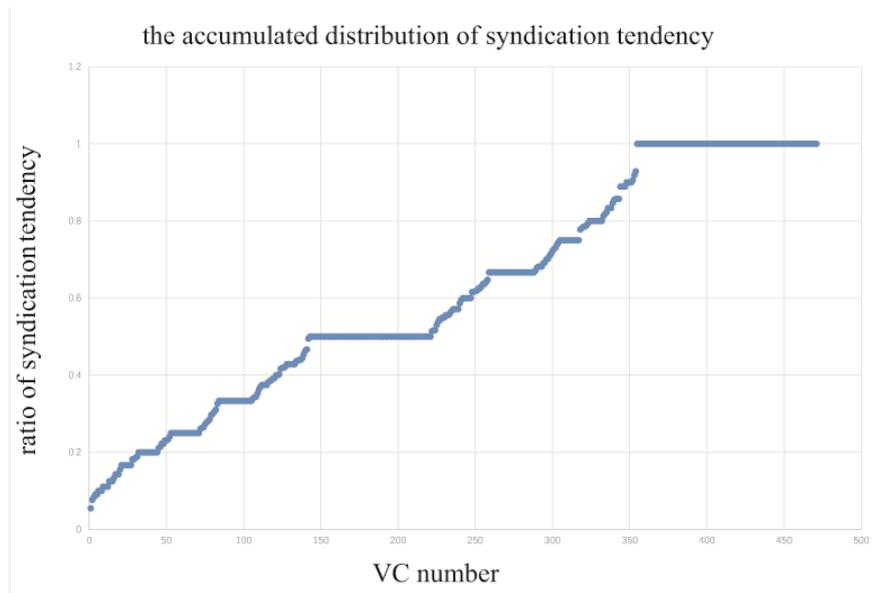

**Fig. 5.** Ranking the VC Firms According to their Syndication Tendency. The X-axis is the ratio of syndication times, and the Y-axis is the accumulative number of VC firms.

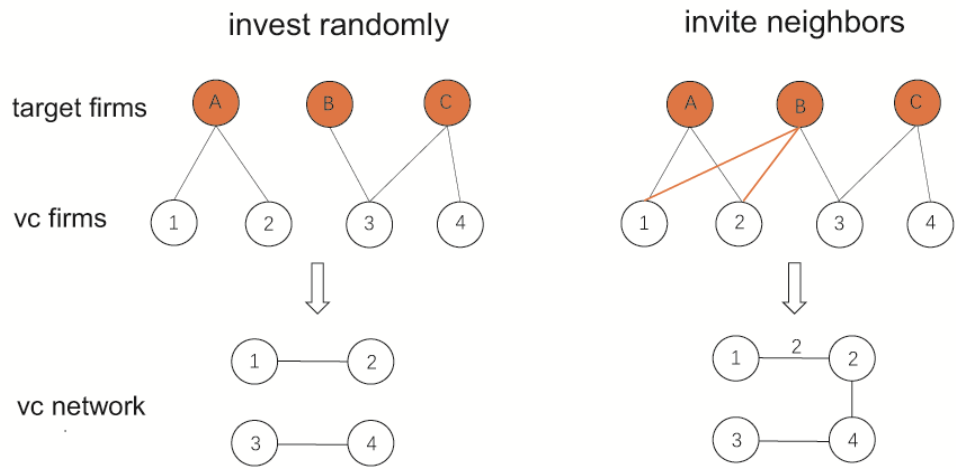

**Fig. 6.** Network Formation in the Relational Embeddedness Model. There are investing and inviting, stages in a time stamp. During the investing stage, a VC firm invests in targeted firms randomly according to its investment frequency. In the inviting step, this VC invites its syndication partners according to its syndication tendency and following the selection mechanism state in Function (3).

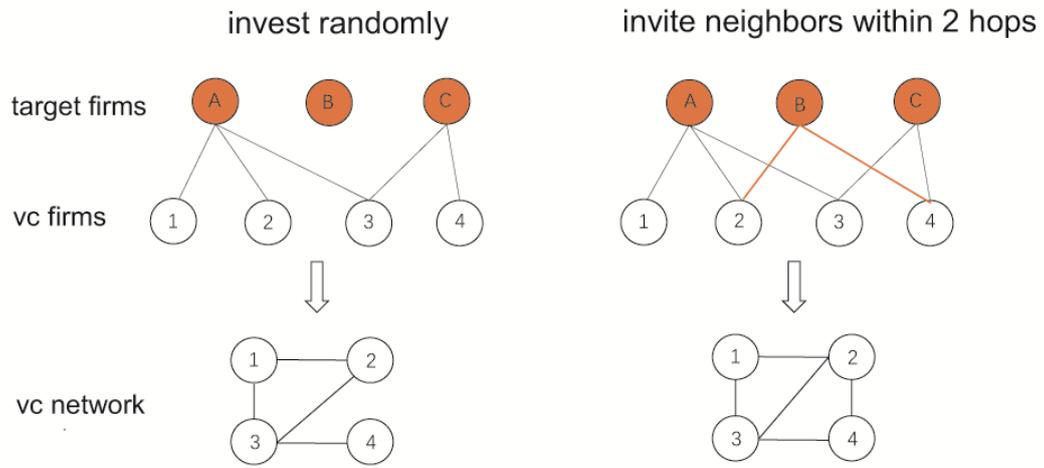

**Fig. 7.** Network Formation in the Structural Embeddedness Model: Based on the relational embeddedness Model, the selection mechanism is changed to Function (4).

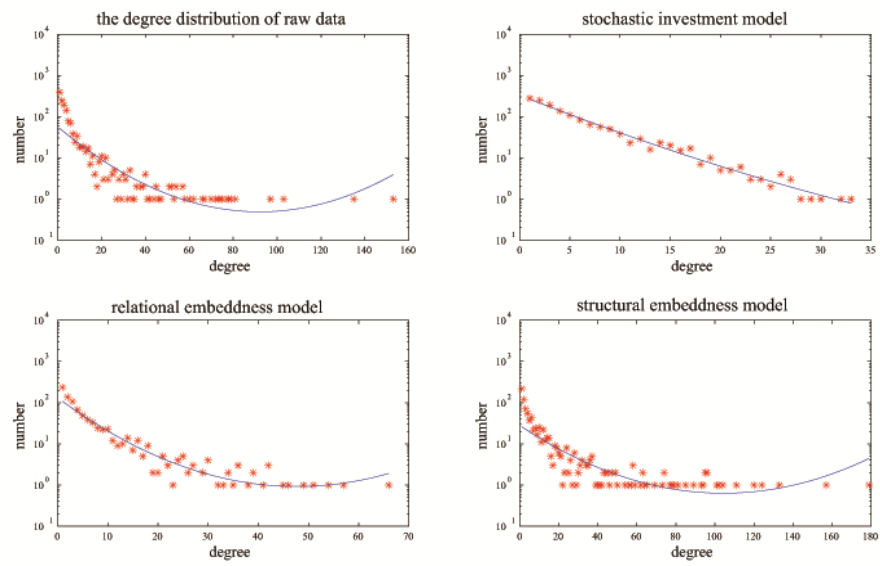

**Fig. 8.** The Comparison of Degree Distribution of Different Models and the Real Network: X-axis is the number of nodes, and Y-axis is the log of node degree.

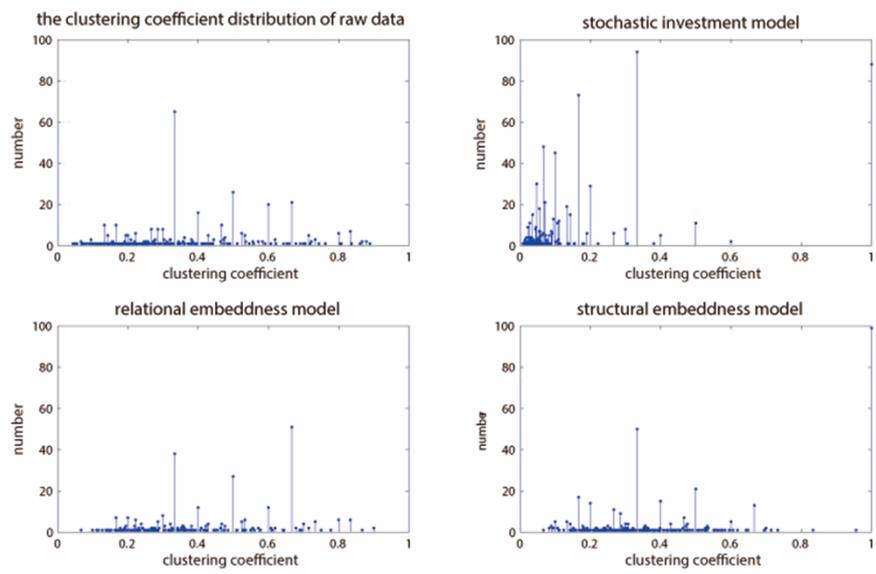

**Fig. 9.** The Comparison of Clustering Coefficient Distribution of Different Models and the Real Network.

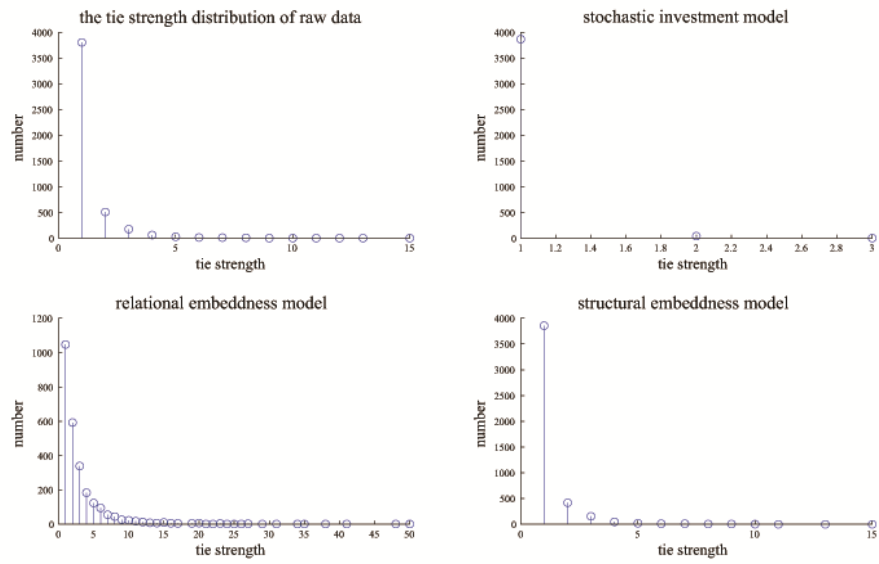

**Fig. 10.** The Comparison of Tie Strength Distribution of Different Models and the Real Network

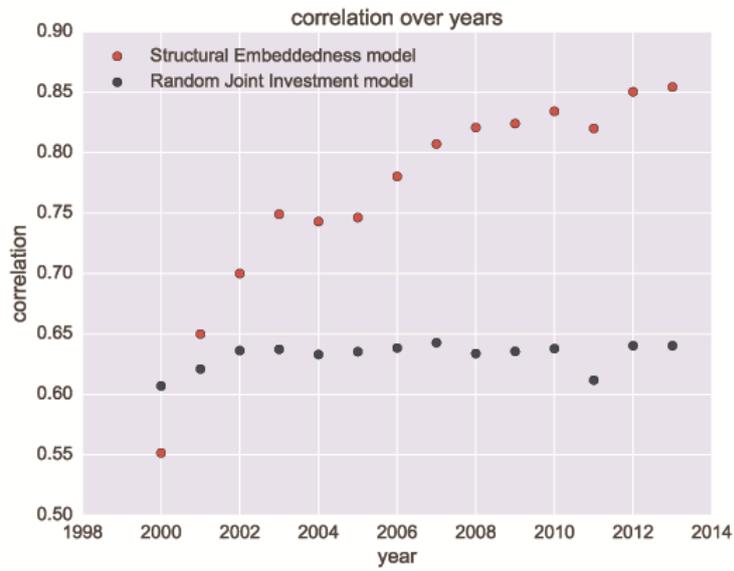

**Fig. 11.** Correlation between the Indicators of Small-World Network with An Elite-Clique: There are 8 indicators for both elites and followers in Table 1, and we correlate the models' indicators with those in the real network. The correlation coefficients between the real network and the structural embeddedness are in red dots. The coefficients between the real network and Random Joint Investment model are in blue dots.

| Network Name | Real Network Property | | | Structural Embeddedness | | Random Investment | |
|:---:|:---:|:---:|:---:|:---:|:---:|:---:|:---:|
| nodes category | elite | follower | all nodes | elite | follower | elite | follower |
| degree | 18.216 | 3.650 | 6.440 | 12.397 | 2.583 | 0.765 | 0.765 |
| k-shell | 12.000 | 2.330 | 3.678 | 7.516 | 1.018 | 0.519 | 0.519 |
| betweenness | 0.014 | 0.002 | 0.008 | 0.019 | 0 | 0 | 0 |
| investment frequency | 47.310 | 4.090 | 8.500 | 75.572 | 4.450 | 0.765 | 0.765 |
| elite-clique density | 0.492 | 0.003 | 0.004 | 0.248 | 0.001 | 0 | 0 |
| elite-clique EI index | 123 | 0.750 | - | 124 | 0.500 | 0 | 0 |
| center-satellites group density | 0.100 | - | - | 0.100 | - | 0.341 | - |
| center-satellites groups' EI index | 25 | - | - | 50 | - | 226 | - |

**Table 1.** The Eight Indicators Illustrating Small-World Network with An Elite-Clique: The Comparison between Structural Embeddedness and Stochastic Joint Models and the Real Network.

network size    network density    network growth rate    investment types

| time step (t = 14) | VC number $m_t = f_A(m_{t-1}) = 75 * 1.3^t$ | target firm number $n_t = f_B(n_{t-1}) = 375 * 1.3^t$ | investment frequency (constant property for nodes) |
|---|---|---|---|
| t = 1 | 75 | 375 | [0.26, 0.80, 5.05] |
| t = 2 | 98 | 488 | [0.26, 0.80, 5.05] |
| ... | | | |
| t = 14 | 2,953 | 14,765 | [0.26, 0.80, 5.05] |

**Table 2.** The Control variables and the Evolutionary Process in the Random Joint Investment Model.

| investment frequency / syndication tendency | high(5.05) | middle(0.80) | low(0.26) |
|---|---|---|---|
| strong(0.96) | (high, strong) | (middle, strong) | (low, strong) |
| middle(0.59) | (middle, strong) | (middle, middle) | (low, middle) |
| weak(0.30) | (weak, strong) | (middle, weak) | (low, weak) |

**Table 3.** The Parameters of Investment Frequency and Syndication Tendency in the Three Models. There are nine types of VC nodes with high, middle, low investment frequency and invitation tendency.

| syndication frequency | invitation probability | Syndication frequency in total |
|:---:|:---:|:---:|
| 1 | 0.04 | 27438 |
| 2 | 0.10 | 3664 |
| 3 | 0.15 | 1068 |
| 4 | 0.18 | 434 |
| 5 | 0.21 | 212 |
| 6 | 0.19 | 84 |
| 7 | 0.46 | 26 |
| 8 | 0.26 | 38 |
| 9 | 0.38 | 16 |
| 10 | 0.75 | 4 |

**Table 4.** The invitation probability in the next run of investment grows with the syndication (co-invest) frequency.

**comparing the motif number between different models and the real network**

| motif type | real | stochastic | structural | relational |
|---|---|---|---|---|
| 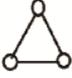 | 7668 | 634 | 7497 | 5435 |
| 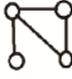 | 633815 | 14642 | 552436 | 7939 |
| 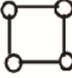 | 0 | 551 | 0 | 0 |
| 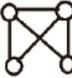 | 58765 | 127 | 53265 | 22518 |
| 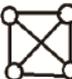 | 10258 | 50 | 9110 | 5912 |

**Table 5.** The Comparison of the Motif Number among Different Models and the Real Network

# Acknowledgements

We acknowledge the financial support of Center for Social Network Research, Tsinghua University and Tsinghua's research project "Mining Tie Strength and Social Capital by Using Telecom Big Data", Project number: 20175080105, as well as the support of Chinese Natural Science Foundation Project "Social Network in Big Data Analysis: A Case in Investment Network", Project number: 71372053.

# References

Ahuja, A. K., Dorn, J. D., Caspi, A., Mcmahon, M. J., Dagnelie, G., Dacruz, L., Greenberg, R. J. (2011). Blind subjects implanted with the Argus II retinal prosthesis are able to improve performance in a spatial-motor task. *British Journal of Ophthalmology*, *95*(4), 539-543.

Axelrod, J., & Reisine, T. D. (1984). Stress hormones: their interaction and regulation. *Science*, *224*(4648), 452-459.

Barabási, A.-L., Jeong, H., Néda, Z., Ravasz, E., Schubert, A., & Vicsek, T. (2002). Evolution of the social network of scientific collaborations. *Physica A: Statistical Mechanics and its Applications*, 311(3), 590-614.

Barabási, A.-L. (2005). The origin of bursts and heavy tails in human dynamics. *Nature*, *435*(7039), 207-211.

Barabási, A.-L., & Albert, R. (1999). Emergence of Scaling in Random Networks. *Science*, *286*(5439), 509-512.

Beckman, C. M., Haunschild, P. R., & Phillips, D. J. (2004). Friends or Strangers? Firm-Specific Uncertainty, Market Uncertainty, and Network Partner Selection. *Organization Science*, *15*(3), 259-275.

Bruton, G. D., & Ahlstrom, D. (2003). An institutional view of China's venture capital industry: Explaining the differences between China and the West. *Journal of Business Venturing*, *18*(2), 233-259.

Bruton, G. D., Fried, V. H., & Manigart, S. (2005). Institutional Influences on the Worldwide Expansion of Venture Capital. *Entrepreneurship Theory and Practice*, *29*(6), 737-760.

Burt, R. S. (1992). *Structural Hole*. Cambridge, MA: Harvard Business School Press.

Burt, R. S. (2000). The Contingent Value of Social Capital. In *Knowledge and Social Capital* (pp. 255-286). Toronto, ON: Elsevier.

Burt, R. S. (2010). *Neighbor Networks: Competitive Advantage Local and Personal*. New York, Oxford: Oxford University Press.

Burt, R. S., & Burzynska, K. (2017). Chinese Entrepreneurs, Social Networks, and Guanxi. *Management and Organization Review*, *13*(2), 221-260.

Burt, R., & Knez, M. (1996). Trust and third-party gossip. In R. Kramer & T. Tyler, *Trust in organizations: Frontiers of theory and research* (pp. 68-89). Thousand Oaks, CA: SAGE Publications.

Burt, R. S., & Opper, S. (2017). Early Network Events in the Later Success of Chinese Entrepreneurs. *Management and Organization Review*, *13*(3), 497-537.

Bygrave, W. D. (1987). Syndicated investments by venture capital firms: a networking perspective. *Journal of Business Venturing*, *2*(2), 139-154.

Garlaschelli, D., & Loffredo, M. I. (2005). Structure and evolution of the world trade network. *Physica A: Statistical Mechanics and its Applications*, *355*(1), 138-144.

Granovetter, M. (1973). The Strength of Weak Ties. *American Journal of Sociology*, *78*(6), 1360-1380.

Granovetter, M. (1985). Economic Action and Social Structure: The Problem of Embeddedness. *American Journal of Sociology*, *91*(3), 481-510.

Granovetter, M. (2017). *Society and Economy: Framework and Principles*. Cambridge, MA: Harvard University Press.

Gu, W., Gong, L., Lou, X., & Zhang, J. (2017). The Hidden Flow Structure and Metric Space of Network Embedding

Algorithms Based on Random Walks. *Scientific reports*, *7*(1), 13114.

Gulati, R. (1999). Network location and learning: the influence of network resources and firm capabilities on alliance formation. *Strategic Management Journal*, *20*(5), 397-420.

Gulati, R., & Westphal, J. D. (1999). Cooperative or Controlling? The Effects of CEO-Board Relations and the Content of Interlocks on the Formation of Joint Ventures. *Administrative Science Quarterly*, *44*(3), 473-506.

Hardin, R. (2001). Conceptions and explanations of trust. *Trust in Society.* New York, NY: Russell Sage Foundation.

Hochberg, Y. V., Ljungqvist, A., & Yang, L. U. (2007). Whom You Know Matters: Venture Capital Networks and Investment Performance. *The Journal of Finance*, *62*(1), 251-301.

Hochberg, Y. V., Ljungqvist, A., & Yang, L. U. (2010). Networking as a Barrier to Entry and the Competitive Supply of Venture Capital. *The Journal of Finance*, *65*(3), 829-859.

Holbrook, M. B. (2002). Complexity and Management: Fad or Radical Challenge to Systems Thinking? *Journal of Macromarketing*, *22*(2), 198-201.

Hwang, K.-k. (1987). Face and Favor: The Chinese Power Game. *American Journal of Sociology*, *92*(4), 944-974.

Jovanovic, B. (1982). Selection and the Evolution of Industry. *Econometrica*, *50*(3), 649-670.

Kingma, D. P., & Ba, J. L. (2015). Adam: Amethod for stochastic optimization. In *International Conference on Learning Representations (ICLR)*. Ithaca, NY: arXiv.org.

Kogut, B., Urso, P., & Walker, G. (2007). Emergent Properties of a New Financial Market: American Venture Capital Syndication, 1960-2005. *Management Science*, *53*(7), 1181-1198.

Lerner, J. (1994). The Syndication of Venture Capital Investments. *Financial Management*, *23*(3), 16-27.

Loftus, B. J., Fung, E., Roncaglia, P., Rowley, D., Amedeo, P., Dan, B., Fraser, J. A. (2005). The Genome of the Basidiomycetous Yeast and Human Pathogen Cryptococcus Neoformans. *Science*, *307*(5713), 1321-1324.

Lou, Y. S., and Li, Y. C. (2017). The emergence of core-peripheral structure in the Chinese VC industry. *Journal of Finance and Accounting* (2): 107-112. (in Chinese)

Luo, J. D., Xiao, H., Burt, R., Chou, C. W., Cheng, M. Y., & Fu, X. M. (2017). Measurement of Guanxi Circles: Using Qualitative Study to Modify Quantitative Measurement: Interdisciplinary Approaches and Case Studies. In *Social Network Analysis* (pp.71-104). Boca Raton: CRC Press.

Luo, J. D., Rong, K., Yang, K., Guo, R., & Zou, Y. (2018). Syndication through social embeddedness: A comparison of foreign, private and state-owned venture capital (VC) firms. *Asia Pacific Journal of Management*, 1-29.

Nguyen, N. P., Dinh, T. N., Tokala, S., & Thai, M. T. (2011). Overlapping Communities in Dynamic Networks: Their Detection and Mobile Applications. In *Proceedings of the 17th Annual International Conference on Mobile Computing and Networking* (pp. 85-96). New York, NY: ACM.

Padgett, J. F., & Powell, W. W. (2012). *The Emergence of Organizations and Markets*. Princeton: Princeton University Press.

Powell, W. W., White, D. R., Koput, K. W., & Owen-Smith, J. (2005). Network Dynamics and Field Evolution: The Growth of Interorganizational Collaboration in the Life Sciences. *American Journal of Sociology*, *110*(4), 1132-1205.

Rosenkopf, L., & Padula, G. (2008). Investigating the Microstructure of Network Evolution: Alliance Formation in the Mobile Communications Industry. *Organization Science*, *19*(5), 669-687.

Small, M. L., & Newman, K. (2001). Urban Poverty after The Truly Disadvantaged: The Rediscovery of the Family, the Neighborhood, and Culture. *Annual Review of Sociology*, *27*(1), 23-45.

Sorenson, O., & Stuart, T. E. (2008). Bringing the Context Back In: Settings and the Search for Syndicate Partners in Venture Capital Investment Networks. *Administrative Science Quarterly*, *53*(2), 266-294.

Tylenda, T., Angelova, R., & Bedathur, S. (2009). Towards Time-aware Link Prediction in Evolving Social Networks. In *Proceedings of the 3rd Workshop on Social Network Mining and Analysis* (p. 9). New York, NY: ACM.

Useem, M. (1984). *The Inner Circle*. New York, NY: Oxford University Press.

Velickovic, P., Cucurull, G., Casanova, A., Romero, A., Lio, P., & Bengio, Y. (2017). Graph attention networks. *arXiv preprint arXiv:1710.10903*.

Viswanath, B., Mislove, A., Cha, M., & Gummadi, K. P. (2009). On the Evolution of User Interaction in Facebook. In *Proceedings of the 2Nd ACM Workshop on Online Social Networks* (pp. 37-42). New York, NY: ACM.

Watts, D. J., & Strogatz, S. H. (1998). Collective dynamics of 'small-world' networks. *Nature*, *393*(6684), 440-442.

Wu, W. F., Wu, C. F., and Rui, M. (2009). Between the special connections that high-ranking managers of some of China's listed companies have with government and tax preference afforded to them. *Management World* (3), 134-142 (in Chinese).

Xiao, Z., & Tsui, A. S. (2007). When brokers may not work: The cultural contingency of social capital in Chinese high-tech firms. *Administrative Science Quarterly* (*52*), 1-31.

Wang, Z., Zhou, Y., Tang, J., & Luo, J. D. (2016). The prediction of venture capital co-investment based on structural balance theory. *IEEE Transactions on Knowledge and Data Engineering*, 28(2), 537-550.